\documentstyle[preprint,aps]{revtex}

\newcommand{\vv}[1]{\mbox{\boldmath{$#1$}}}

\def\xiloc{\xi_{\rm loc}}

\def\wc{W_{\rm c}}
\def\beq{\begin{equation}}
\def\eeq{\end{equation}}

\begin{document}

\title{Anomalous diffusion at the Anderson transitions}

\author{Tomi Ohtsuki}

\address{Department of Physics, Sophia University,
Kioi-cho 7-1, Chiyoda-ku, Tokyo 102, Japan}

\author{Tohru Kawarabayashi}
\address{Institute for Solid State Physics, University of Tokyo,
Roppongi, Minato-ku, Tokyo 106, Japan}

\date{\today}
\maketitle

\begin{abstract}
Diffusion of electrons in three dimensional disordered
systems is investigated numerically for all the three universality classes,
namely, orthogonal, unitary and symplectic ensembles.
The second moment of the wave packet $<\vv{r}^2(t)>$ at the
Anderson transition is shown to behave as $\sim t^a (a\approx 2/3)$.
From the temporal autocorrelation function $C(t)$, the fractal dimension $D_2$ is
deduced, which is almost half the value of space dimension
for all the universality classes.
\end{abstract}

Metal-insulator transitions are one of the most extensively investigated
subjects in condensed matter physics.
Especially interesting is the quantum phase transition, where the
transition is driven by changing the parameter of quantum systems
instead of temperature.
The Anderson transition\cite{Anderson,KM} is a typical example,
where extended electronic states become localized with the
increase of disorder.

Much effort has been devoted to clarify the Anderson transition,
both experimentally and theoretically.
In the metallic regime where the electronic states are extended,
the transition is defined by the vanishing conductivity $\sigma$
as the strength of disorder $W$ is increased.
It is characterized by the critical exponent $s$ as
\beq
\sigma \sim (\wc -W)^{s} ,
\eeq
with $\wc$ the critical disorder.
In the insulating regime where the states are localized,
it is most clearly seen by the divergence of the localization length
$\xiloc$ as
\beq
\xiloc\sim (W-\wc)^{-\nu} .
\eeq
From the one parameter scaling theory,\cite{gang,kaw} the exponent $\nu$ 
is related to $s$ by the Wegner's scaling law, \cite{weg}
\beq
s=(d-2)\nu ,
\eeq
$d$ being the dimensionality of the system, and once $\nu$ is
determined, we can predict the behavior of the conductivity near
the Anderson transition.

The behaviors of the localization length and the conductivity
are conjectured to be universal, i.e., $\nu$ does not
depend on the detail of the system.
They are determined only by the basic symmetry of the system under
the operation of time reversal or spin rotation.\cite{weg,HLN}
Systems with time reversal and spin rotation symmetry are
called orthogonal ensemble, while those with only time reversal
symmetry are called symplectic ensemble.
Systems without time reversal symmetry are unitary
ensemble.

The value of $\nu$ has been calculated for three dimensional
system by using the finite-size scaling argument.
It is estimated to be $1.4\pm 0.1$ for orthogonal ensemble,\cite{KM}
$1.3\pm 0.2$ for unitary ensemble,\cite{OKO,HKO,CD}
and again $1.3\pm 0.2$ for symplectic ensemble.\cite{KOSO,hof}
These facts indicate that the critical behavior of
conductivity as well as the localization length does not depend
significantly on the symmetry of the system.

On the other hand, recent analyses on the energy level statistics at
the Anderson transition show that the level statistics
do depend on the symmetry of the system, though they are
independent of system size or
model.\cite{KOSO,hof,SSSLS,HS_3DO,OO_0,ZK,SZ,OO,Evangelou_2DS,HS,BSZK}
This universal behavior is related to the scale invariance
at the transition, where eigenfunctions show fractal structure.
Peculiar behavior just at the transition is now attracting
a lot of attention.

In this paper, we numerically discuss electron diffusion at the Anderson
transition in three-dimensional (3D) disordered systems.
The diffusion coefficient becomes size dependent at the transition,
which leads to the increase of
the mean square diffusion length $<\vv{r}^2(t)>$
as $t^a$ with the exponent $a<1$. \cite{Imry}
The return probability also decreases as power law, reflecting
the fractal structure of the wave function.\cite{cha,BHS}

\medskip 

First we discuss the behavior of $<\vv{r}^2(t)>$.
As in the percolation theory,\cite{stau} let us assume the scaling form
\beq
<\vv{r}^2(t)> = C t^{k_1} f\left(
(\wc -W)t^{k_2}
\right)  . \label{eq:anomr}
\eeq
In the metallic regime, $<\vv{r}^2(t)>$ increases as
$2dDt$ where $D$ is the diffusion constant.
It is related to the conductivity from the Einstein relation, 
and behaves as $(\wc -W)^s$.
Therefore, $f(x)$ should be proportional to $x^s$ in the limit of
large $x$.

In the insulating regime, the wave packet ceases to diffuse if the
diffusion radius becomes the order of localization length.
Therefore we have $<\vv{r}^2(t)>\sim \xiloc^2\sim (\wc -W)^{-2\nu}$
and $f(x)$ is proportional to $(-x)^{-2\nu}$ when $-x$ is large enough.
From this argument, we have
\beq
\begin{array}{ll}
k_1+s k_2 &=1 {\ },\\
k_1-2\nu k_2 &=0 {\ },
\end{array}
\eeq
and consequently
\beq
k_1=\frac{2\nu}{s+2\nu} , {\ \ } k_2 = \frac{1}{s+2\nu}  .
\eeq
Using the scaling relation $s=(d-2)\nu$, we have
\beq
k_1=\frac{2}{d} , {\ \ } k_2 = \frac{1}{d\nu}  .
\eeq
We therefore expect
\beq
<\vv{r}^2(t)>\sim t^{2/d} , \label{eq:anomr2}
\eeq
at the Anderson transition $(W=W_{\rm c})$.

At the transition, the wave function has fractal structure.
In this situation, if the radius of the wave packet is $r$,
the return probability $P(t)$ is proportional to $r^{-D_2}$.
From (\ref{eq:anomr2}), we have
\beq
P(t) \sim t^{-D_2/d} . \label{eq:anomret}
\eeq
From the behavior of $P(t)$, we can determine the fractal dimension $D_2$.
This intuitive argument agrees with the more detailed analysis of the
scaling behavior of the dynamical diffusion coefficient
$D(\vv{q},\omega)$.\cite{cha,BHS}

\medskip                                                                        
In the actual simulation, we have adopted tight binding Hamiltonian
\beq
 H = \sum_{<i,j>,\sigma,\sigma'} 
 V_{i,\sigma;j,\sigma'}c^{\dagger}_{i,\sigma}
 c_{j,\sigma'} + \sum_{i,\sigma} W_ic^{\dagger}_{i,\sigma}
 c_{i,\sigma} ,
\eeq
where $i, j$ denote the lattice site, and $\sigma, \sigma'$ the spin.
In the orthogonal case, $ V_{i,\sigma;j,\sigma'}=V\delta_{\sigma,\sigma'}$
is real,while
$ V_{i,\sigma;j,\sigma'}$ is
$V \exp (i\phi_{i,j})\delta_{\sigma,\sigma'}$
with  $\phi_{i,j}$ the Peierls phase factor in the unitary case.
 In both cases, no spin flip process is included.
In the symplectic case, the hopping is described by
\begin{equation}
 V_{i,\sigma;i- k,\sigma'} =  V [\exp( -{\rm i}\theta 
 \sigma_{k}) ]_{\sigma,\sigma'}, \qquad k = \hat{x}, \hat{y}
 , \hat{z}, \label{coupling}
\end{equation}
where $\sigma_k$'s are  Pauli matrices.\cite{KOSO,keith}
We have assumed the simple cubic structure with the lattice constant
taken to be unity.
Only the nearest neighbor coupling is assumed.
The site-diagonal potentials $W_i$ are assumed to be 
distributed independently, and their distribution 
is taken to be uniform in the range $[-W/2,W/2]$.

Instead of diagonalizing the system directly,
we solve numerically the time-dependent Schr\"{o}dinger
equations.\cite{KO1,KO2}
We evaluate the time-evolution operator 
$U(t) = \exp(-{\rm i}Ht/\hbar)$ by using 
the decomposition formula for exponential operators.\cite{Suzuki} 
The $n$-th order decomposition $U_n$
satisfies the condition
\begin{equation}
U(\delta t) = U_n(\delta t) + O(\delta t^{n+1}) . \label{a1}
\end{equation}
We have adopted the same forth-order decomposition,
as in the previous papers,\cite{KO1,KO2}  given by
\begin{equation}
U_4 = U_2(-{\rm i}pt/\hbar)U_2(-{\rm i}(1-2p)t/\hbar)U_2(-{\rm i}pt/\hbar) 
\label{ourdec}
\end{equation}
with
$$
U_2(x) \equiv 
{\rm e}^{xH_1/2}\cdots{\rm e}^{xH_{q-1}/2}{\rm e}^{xH_q}
{\rm e}^{xH_{q-1}/2}\cdots{\rm e}^{xH_1/2}, $$
where $H=H_1 + \cdots + H_q$ and $p=(2-\sqrt[3]{3})^{-1}$. 
The decomposition is made so that each Hamiltonian $H_i (i=1,\cdots , q)$
should consist of commuting terms.

The actual simulations have been done in systems with
$59\times59\times59$ lattice sites
for orthogonal and unitary ensembles, while $69\times69\times69$
lattice sites are used to discuss the symplectic case.
In each case, average over 10 independent realizations of
random potentials has been
performed.
The initial wave packet is build by diagonalizing a spherical system with
radius $R=3$ located at the center of the whole system.
We use the wave packet whose energy is closest to the band center.

To discuss the properties at the vicinity of the transition,
we have set $W=\wc=16.5 V$ for orthogonal case. \cite{KM}
In the unitary case we assume that the magnetic field is parallel to
the $z$-direction, and the magnetic flux penetrating the $x$-$y$ plane
unit cell is set to be 0.1 times the flux quantum.
The resulting critical disorder is $W=\wc=17.8 V$.\cite{HKO}
For the symplectic ensemble, we have set $\theta=\pi/6$ in (\ref{coupling})
and $W$ is again set to the critical value $\wc=19.0 V$.\cite{KOSO}
The time step is chosen to be $\delta t = 0.2 \hbar/V_1$ where
$V_1$ is the hopping amplitude without spin flip process.

We first discuss the critical behavior of the second moment of the wave packet
$<\vv{r}^2(t)>$ defined as
\beq
<\vv{r}^2(t)>_{\rm c}\equiv <t|\vv{r}^2|t>-
<t|x|t>^2-<t|y|t>^2-<t|z|t>^2 ,
\eeq
where $|t>$ denotes the state at time $t$.
In Fig. 1, we plot it
as the function of time $t V/\hbar$
for the three universality classes.
The solid line corresponds to the orthogonal case,
the broken line to the unitary, and the dotted line to the symplectic one.
The standard deviations
  with respect to 10 realizations of random potential 
   configurations are typically less than 5$\%$.
From these behavior, $<\vv{r}^2(t)>_{\rm c}$ is estimated to increase as
$t^a$, with $a = 0.67\pm 0.02$ for orthogonal case,
$0.66\pm 0.02$ for unitary case and $0.69\pm 0.02$ for
symplectic case.
The excellent agreement of $a$ with $2/d=2/3$ confirms 
the scaling form (\ref{eq:anomr}) and
the scaling relation $s=(d-2)\nu$.

\bigskip 
\centerline{\bf [ Fig. 1 ]} 
\bigskip  

Now we discuss the return probability.
In Fig.2, we plot the temporal autocorrelation function\cite{KPG} $C(t)$
defined by the overlap function between the initial state and
the state at time $t$ as
\beq
C(t)\equiv \frac{1}{t} \int_0^t {\rm d}t' |<t'|0>|^2 
=\frac{1}{t} \int_0^t {\rm d}t' P(t') .
\eeq
Average of $\log C(t)$
over 10 random potential configurations has been performed,
and the standard deviation is indicated for orthogonal case
(the standard deviations for the
other cases are not shown but they are almost the same).
From the asymptotic behavior, we estimate the fractal dimension
$D_2$ as $1.5\pm 0.2$ for orthogonal, $1.7\pm 0.2$ for unitary
and $1.6\pm 0.2$ for symplectic case.
In the case of orthogonal ensemble, the value $1.5\pm 0.2$ agrees with
the results obtained previously by the direct
diagonalization.\cite{BHS,SE,SG}
Our new results for unitary and symplectic cases show that
the fractal dimensionality $D_2$ does not depend  strongly on the
symmetry.
The results are summarized in Table I.

\bigskip 
\centerline{\bf [ Fig. 2]} 
\bigskip

In conclusion, we have studied the diffusion of
electron in 3D disordered systems
at the Anderson transition by numerically solving the
Schr\"{o}dinger equation.
The anomalous diffusion $\vv{r}^2(t)\sim t^{2/3}$ has been clearly
observed, which is expected from the scaling form (\ref{eq:anomr})
and the scaling relation $s=(d-2)\nu.$
The fractal dimensionality $D_2$ is also estimated.
For all three universality classes, $D_2$ is almost half the
space dimension.
It is interesting to note that the values
$D_2$ for two dimensional Anderson
transitions, namely the quantum Hall and symplectic systems, are
almost the same as well ($1.62\pm 0.02$ for the former \cite{aoki,HKS}
and $1.66\pm 0.05$ for the latter.\cite{KO2,Schweitzer}).
As discussed by Brandes {\it et al.},\cite{BHS,BSK} the fractal properties are
reflected in the temperature dependence of the inelastic scattering time
$\tau_{\rm in}$
at the Anderson transition in relatively high temperature.
Our results for unitary and symplectic cases indicate
that almost the same temperature dependencies are observed
in all 3D Anderson transitions.

Finally, let us discuss the temperature dependence of the
conductivity at the transition.
By the anomalous diffusion, the relation between the inelastic
scattering length $l_{\rm in}$ and the inelastic scattering time
is modified to be
\beq
l_{\rm in} \propto  \tau_{\rm in}^{1/3} .
\eeq
Then the effective diffusion constant $D_{\rm eff}$ observed at
finite temperatures
is 
\beq
D_{\rm eff} \sim \frac{l_{\rm in}^2}{\tau_{\rm in}}\sim \tau_{\rm in}^{-1/3}  ,
\eeq
leading to $\sigma\sim \tau_{\rm in}^{-1/3}$.
At sufficiently low temperature, $\tau_{\rm in}^{-1}$ is proportional
to the temperature $T$.
Experimentally observed $T^{1/3}$ behavior of the conductivity at the
transition is thus consistent with the present scaling argument
using the scaling relation $s=(d-2)\nu$.
It is recently suggested \cite{BK} that the scaling relation
is modified in the interacting system where Anderson-Mott transition
occurs.
Careful investigation of the temperature dependence of the conductivity
at the transition will clarify the nature of the transition.

\noindent                                                                       
                                                                                
\medskip                                                                        
The authors are grateful to Professor Yoshiyuki Ono and
Dr. T. Brandes for fruitful discussions.
This work is in part financed by the
Grants-in-Aid 08740327
from the Ministry of Education, Science and Culture. 
 The numerical calculations have been
in part performed on a FACOM VPP500 of 
Institute for Solid State Physics, University of Tokyo. 
                                                                              
\vfil\eject                                                                     
%
%

\def\pr{Phys. Rev. }                                                            
\def\prl{Phys. Rev. Lett. }                                                     
\def\jpsj{J. Phys. Soc. Jpn. }                                                  
\def\ssc{Solid State Commun. }

\vfill\eject

\begin{table}
\caption{
Summary of the exponent $a$ and the fractal dimensionality $D_2$
for three universality classes.}
\label{table:1}
\begin{tabular}{@{\hspace{\tabcolsep}\extracolsep{\fill}}cccc} \hline
{\ \ }          & orthogonal        &  unitary             &   symplectic    \\ \hline
$a$            &$0.67\pm 0.2$  &$0.66\pm 0.2$    & $0.69\pm 0.2$  \\ \hline
$D_2$        &$1.5\pm 0.2$    &$1.7\pm 0.2$      & $1.6\pm 0.2$   \\ \hline
\end{tabular}
\end{table}

\bigskip

{\bf Figure captions}

\medskip

Fig. 1: The growth of the second moment $ < r^2 (t) >_{\rm c}$ 
       of the wave packet.  The solid line corresponds to the
       orthogonal case, the broken line to the unitary and the
       dotted line to the symplectic one.
In large $t$ regime, $t^{2/3}$ behavior is clearly seen.

\medskip

Fig. 2: The time-dependence of the auto-correlation function $C(t)$ .
The bars around  the data for orthogonal case
  indicate the standard deviation
  with respect to 10 realizations of random potential 
   configurations.
They are almost the same for all universality classes, so
only those
 for the orthogonal case are shown for simplicity.
 
\end{document}